\documentclass[twocolumn,showpacs,preprintnumbers,amsmath,amssymb]{revtex4}

\usepackage{graphicx}
\usepackage{dcolumn}
\usepackage{bm}

\begin{document}

\preprint{}

\title{Spin-dependent edge-channel transport in a Si/SiGe quantum Hall system}

\author{K. Hamaya,$^{1, a)}$ S. Masubuchi,$^{1}$ K. Hirakawa,$^{1}$ S. Ishida,$^{2}$ Y. Arakawa,$^{2}$ K. Sawano,$^{3}$ Y. Shiraki,$^{3}$ and T. Machida$^{1,}$$^{4, b)}$}%
\affiliation{%
 $^{1}$Institute of Industrial Science, The University of Tokyo,  4-6-1 Komaba, Meguro-ku, Tokyo 153-8505, Japan.
\\$^{2}$Nanoelectronics Collaborative Research Center (NCRC), IIS and RCAST, The University of Tokyo, 4-6-1 Komaba, Meguro-ku, Tokyo 153-8505, Japan. 
\\$^{3}$Research Center for Silicon Nano-Science, Advanced Research Laboratories, Musashi Institute of Technology, 8-15-1 Todoroki, Setagaya-ku, Tokyo 158-0082, Japan.
\\$^{4}$Nanostructure and Material Property, PRESTO, Japan Science and Technology Agency, 4-1-8 Honcho, Kawaguchi 332-0012, Japan.
 }%


\date{\today}

\begin{abstract}
We study the edge-channel transport of electrons in a high-mobility Si/SiGe two-dimensional electron system in the quantum Hall regime. By selectively populating the spin-resolved edge channels, we observe suppression of the scattering between two edge channels with spin-up and spin-down. In contrast, when the Zeeman splitting of the spin-resolved levels is enlarged with tilting magnetic field direction, the spin orientations of both the relevant edge channels are switched to spin-down, and the inter-edge-channel scattering is strongly promoted. The evident spin dependence of the adiabatic edge-channel transport is an individual feature in silicon-based two-dimensional electron systems, originating from a weak spin-orbit interaction. 
 
\vspace{5mm}$^{a)}$ E-mail:  hamaya@iis.u-tokyo.ac.jp

$^{b)}$ E-mail:  tmachida@iis.u-tokyo.ac.jp 
\end{abstract}

\maketitle


Characteristic features such as the valley degree of freedom,\cite{Ando} a metal$-$insulator transition at zero field,\cite{Kravchenko} and a significant anisotropy of magnetotransport properties in the quantum Hall regime\cite{Zeitler} have been discovered for silicon-based two-dimensional electron gas (2DEG) systems in silicon metal-oxide-semiconductor field-effect transistors (MOSFETs) and Si/SiGe heterostructures. For these systems, the quantum Hall (QH) effects and their related physics at the Landau-level crossing which is so-called coincidence have been explored in tilted magnetic fields so far.\cite{Zeitler,Zeitler2,Weitz,Schumacher,Koester} Recently, the manipulation of the valley degree of freedom by changing the gate-bias voltage to tune the coincidence condition was further exploited in SiO$_{2}$/Si/SiO$_{2}$ quantum wells.\cite{Takashina} 

In a single-particle picture, the Zeeman splitting ($\Delta E_\mathrm{z}$) depends on the total magnetic field ($B$$_\mathrm{total}$) while the cyclotron energy, $\hbar$$\omega$$_\mathrm{c}$, depends on the perpendicular component ($B$$_\mathrm{\bot}$) of $B$$_\mathrm{total}$. Thus, when we apply the parallel component ($B$$_\mathrm{//}$) in addition to the $B$$_\mathrm{\bot}$ with tilting an external magnetic field direction $\theta$ between the direction of an applied magnetic field and the direction normal to the 2DEG plane, the $\Delta E_\mathrm{z}$ of the spin-resolved levels can be enlarged, giving rise to a crossover of the Landau levels at a certain $\theta$ as schematically illustrated in Fig. 1(a). Using the tilted magnetic fields, we can determine the effective $g$-factor ($g$*),\cite{Weitz,Schumacher,Koester} and one deduces that the value of  $g$* is concerned with carrier density for Si/SiGe heterostructures.\cite{Koester} Also, it was indicated that an exchange interaction between different Landau levels is enhanced under the coincidence condition, showing an overshoot of the Hall resistance at the filling factor of $\nu =$ 3 \cite{Weitz} and transition peaks with unexpectedly huge resistance in the Shubnikov-de Haas (SdH) oscillations.\cite{Schumacher,Koester,Zeitler2} 

Though the edge-channel picture is crucial to understand the electronic transport in QH systems,\cite{Buttiker,Komiyama} few studies of the edge-channel transport have been reported for the silicon-based 2DEG systems. More than ten years ago, a preliminary work using Si-MOSFETs with mobility below 2.0 m$^{2}$/Vs was demonstrated,\cite{Son} but a collective view of the edge-channel transport has not been established because of the low mobility of Si-MOSFETs. Owing to development of high-quality Si/SiGe heterostructures,\cite{Growth} however, the mobility value increases up to $\sim$ 50 m$^{2}$/Vs,\cite{Growth,Weitz} in consequence, the fractional QH effect can be explored \cite{Lai} and a possibility of spin-based quantum computing applications was indicated.\cite{Tyryshkin} Using these high-quality Si/SiGe heterostructures, we can elucidate the edge-channel transport controlled by tuning the coincidence condition: at the filling factor of $\nu =$ 4, the edge channels with spin-down 0 $\downarrow$ and spin-up 0 $\uparrow$ are presented in $\hbar$$\omega$$_\mathrm{c}$ $>$ $\Delta E_\mathrm{z}$ while the edge channels with spin-down 0 $\downarrow$ and spin-down 1 $\downarrow$ are formed in $\hbar$$\omega$$_\mathrm{c}$ $<$ $\Delta E_\mathrm{z}$, as shown in Fig. 1(b).
\begin{figure}
\includegraphics[width=8.5cm]{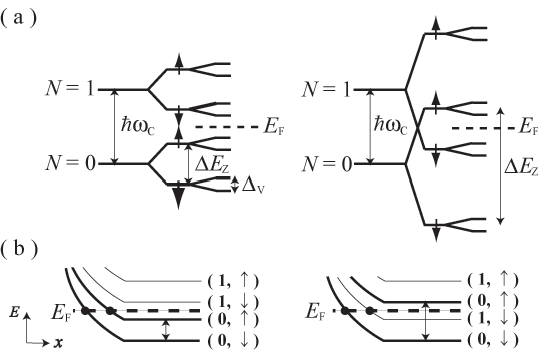}
\caption{( a ) Energy diagrams of Landau levels between $N = 0$ and $N = 1$ for $\hbar$$\omega$$_\mathrm{c}$ $>$ $\Delta E_\mathrm{z}$ (left) and $\hbar$$\omega$$_\mathrm{c}$ $<$ $\Delta E_\mathrm{z}$ (right). The Fermi level $E_\mathrm{F}$ is located at the filling factor of $\nu_\mathrm{}$ = 4. The valley splittings ($\Delta_\mathrm{V}$) are also depicted. ( b ) Edge channel dispersions for $\hbar$$\omega$$_\mathrm{c}$ $>$ $\Delta E_\mathrm{z}$ (left) and $\hbar$$\omega$$_\mathrm{c}$ $<$ $\Delta E_\mathrm{z}$ (right) for $\nu_\mathrm{}$ = 4 (a 2-channel case). The spin orientation of the relevant edge channels switches from (0$\downarrow$, 0$\uparrow$) to (0$\downarrow$, 1$\downarrow$) through the coincidence angle.}
\end{figure} 

In this paper, we report on the observation of the spin-dependent edge-channel transport in a high-mobility Si/SiGe heterostructure in the QH regime. By selectively populating the spin-resolved edge channels, the Hall resistance deviates largely from the quantized value, indicating the first observation of the adiabatic edge-channel transport of electrons in the Si/SiGe heterostructure. The inter-edge-channel (IEC) scattering is strongly suppressed over macroscopic distance between (0 $\downarrow$, 0 $\uparrow$) edge channels while that is significantly promoted between (0 $\downarrow$, 1$\downarrow$) edge channels. The spin dependence clearly observed is a characteristic property of silicon-based QH systems, being due to a small contribution of the spin-orbit interaction to the spin-flip IEC scattering. 
\begin{figure}[b]
\includegraphics[width=8.5cm]{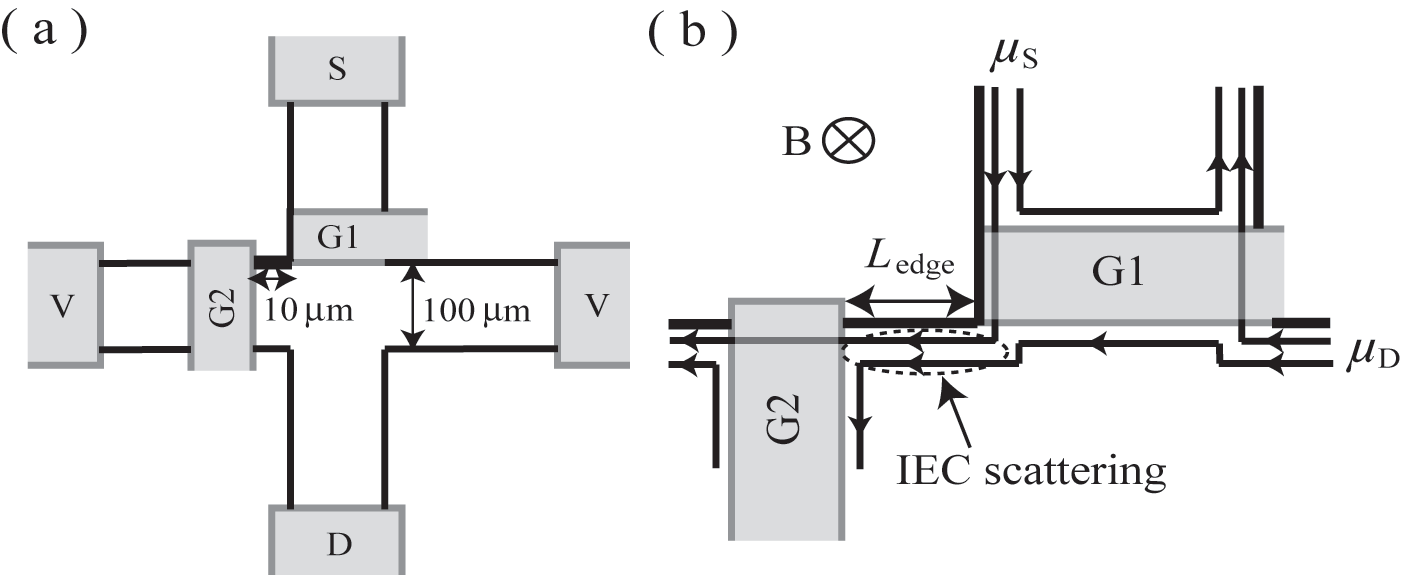}
\caption{( a ) A schematic illustration of the Hall-bar sample. ( b ) The enlarged figure of $L$$_\mathrm{edge}$ region for a 2-channel case. The arrows indicate the direction of electron drift in the edge channels.}
\end{figure}

A high-mobility Si/Si$_{0.75}$Ge$_{0.25}$ heterostructure studied was grown by molecular beam epitaxy (MBE) on the strained-relaxed Si$_{0.75}$Ge$_{0.25}$ buffer layer smoothed by chemical mechanical polishing (CMP).\cite{Sawano} The wafer has the electron mobility of 20 m$^{2}$/Vs and the electron density of  1.35 $\times$ 10$^{15}$ m$^{-2}$ at 0.3 K. For transport measurements, the wafer was patterned into 100-$\mu$m-wide Hall bars with four alloyed AuSb ohmic contacts and two front gates (G1 and G2) crossing the channel as depicted in Fig. 2(a). The front gate structure is composed of a 100-nm-thick SiO$_{2}$ insulating layer grown by plasma enhanced chemical vapor deposition (PECVD) below 400$^{\circ}$C, followed by 2.5-nm-thick Ti/200-nm-thick Au layer deposited by electron-beam evaporation. The distance of the edge region in the Hall bar between the two gates is $L$$_\mathrm{edge}  =$ 10 $\mu$m. The filling factors of Landau levels in the bulk region and under the front gate, $\nu_\mathrm{B}$ and $\nu_\mathrm{G}$, are controlled by adjusting the magnetic field $B$ and the gate-bias voltage $V_\mathrm{G}$. Transport measurements were basically performed using standard lock-in techniques (18 Hz) with an alternating current of 1.0 nA in a $^{3}$He$-$$^{4}$He dilution refrigerator. The SdH oscillations were observed evidently and the longitudinal resistance ($R_{xx}$) showed the plateau corresponding to zero resistance at $\nu =$ 1, 2, and 4. 
\begin{figure}
\includegraphics[width=8cm]{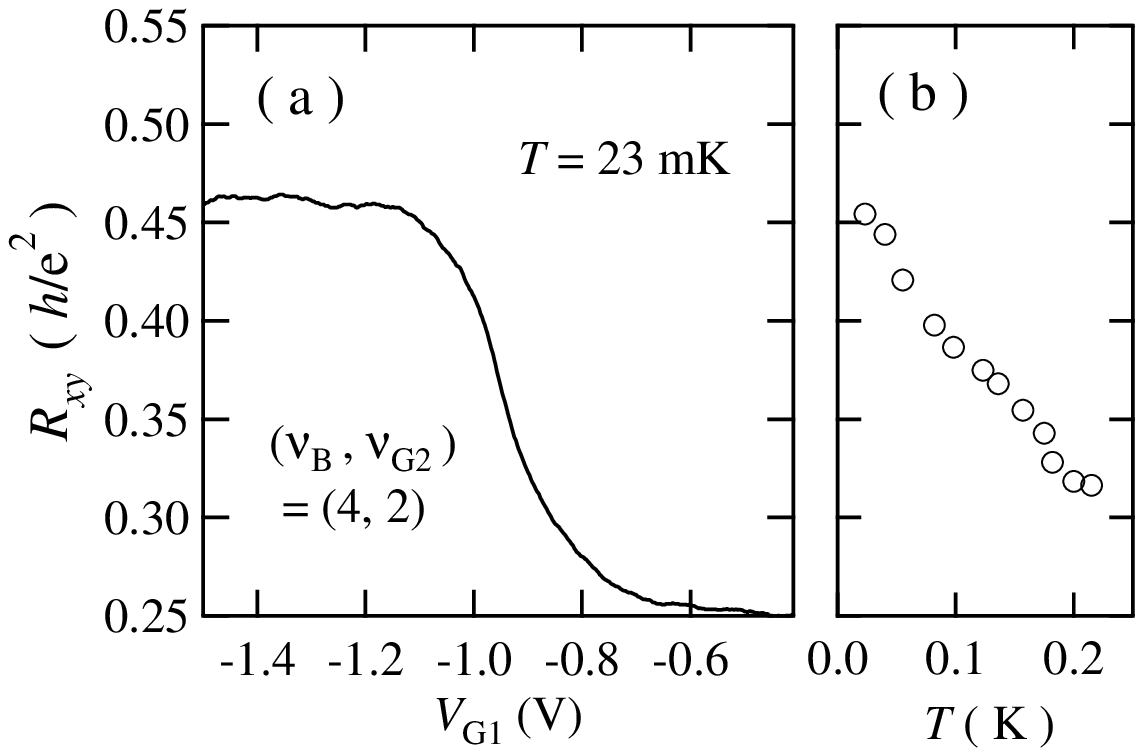}[t]
\caption{$R_{xy}$ as a function of $V_\mathrm{G1}$ at ($\nu_\mathrm{B}$, $\nu_\mathrm{G2}$) = (4, 2) at 23 mK. (b) Temperature dependence of $R_{xy}$ at ($\nu_\mathrm{B}$, $\nu_\mathrm{G1}$, $\nu_\mathrm{G2}$) = (4, 2, 2).}
\end{figure}
 
To examine the edge-channel transport, we focus on the IEC scattering for a 2-channel case as shown in Fig. 2(b).\cite{ref} We hereafter define the electrochemical potentials of the source and the drain reservoirs as $\mu_\mathrm{S}$ and $\mu_\mathrm{D}$, respectively. When $\nu_\mathrm{B} =$ 4 and $\nu_\mathrm{G} =$ 2, the outer channel passes through the two front gates (G1 and G2) while the inner channel is reflected by the gates. Here, the value of $V_\mathrm{G}$ for $\nu_\mathrm{G} =$ 2 was determined experimentally by the measurements of $R_{xx}$ $vs$ $V_\mathrm{G}$.\cite{Komiyama,Komiyama2,Hirai,Muller,Muller2} As a consequence, the electrochemical potential of the outer channel ($\mu_\mathrm{S}$) is different from that of the inner channel ($\mu_\mathrm{D}$) at $L$$_\mathrm{edge}$ shown in Fig. 2(b). For 2DEG in AlGaAs/GaAs heterostructures, many experimental and theoretical studies of the edge-channel transport have been reported,\cite{Komiyama,Komiyama2,Komiyama3,Hirai,Muller,Muller2,Khaetskii,Wees,Alphenaar,Takagaki} in which the IEC scattering is suppressed over macroscopic distance, resulting in a deviation of the Hall resistance ($R_{xy}$) from the quantized value at the QH regime. On the basis of the Landauer-B\"uttiker formalism,\cite{Buttiker2} the adiabatic transport in spin-resolved edge channels at $\nu_\mathrm{B} =$ 4 is likely to indicate $R_{xy} =$ $h$/2$e$$^{2}$\cite{Komiyama3} while the non-adiabatic edge-channel transport shows the quantized value $h$/4$e$$^{2}$ in the case of 2DEG in Si/SiGe heterostructures. 
\begin{figure*}
\includegraphics[width=15cm]{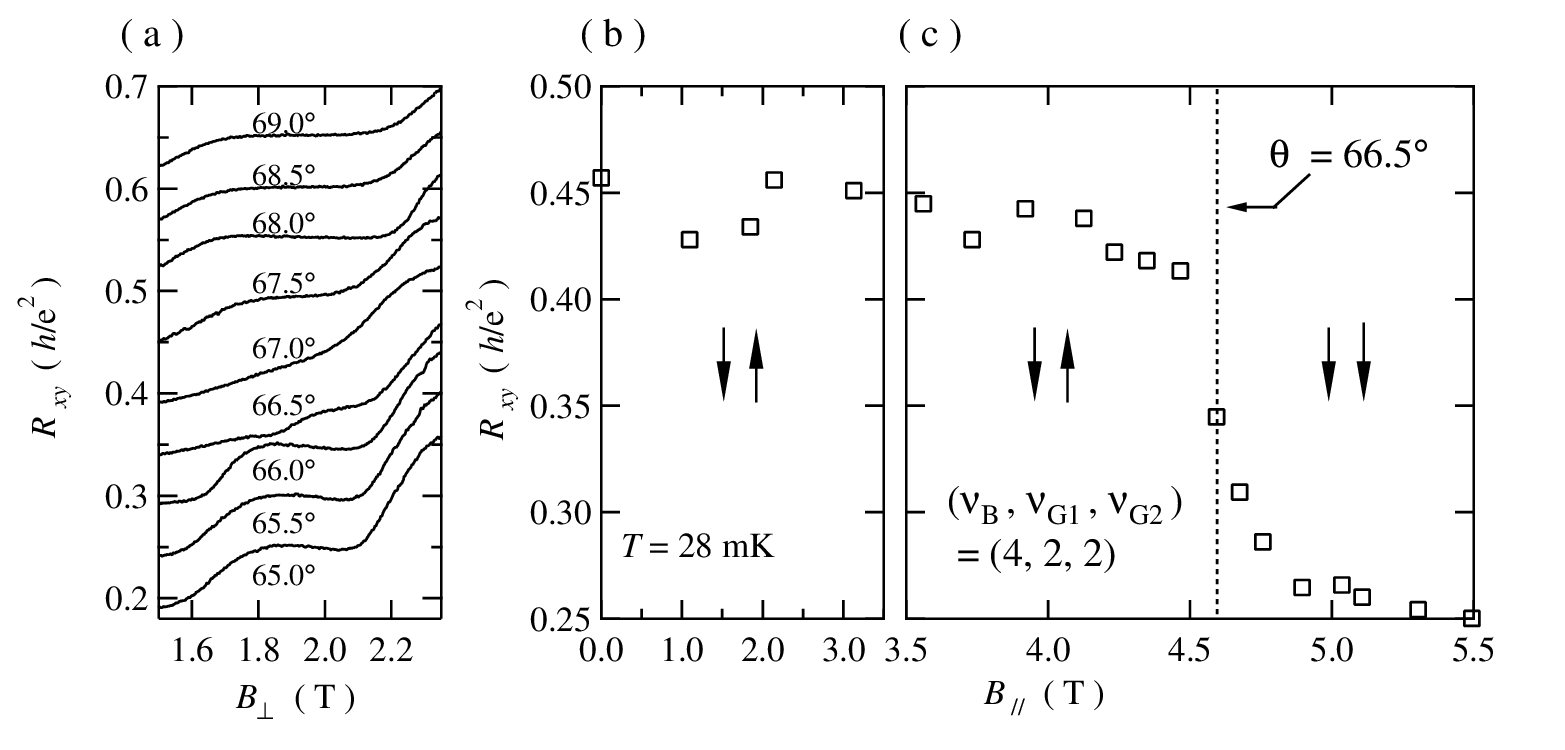}
\caption{( a ) Plots of $R_{xy}$ $vs$ $B$$_\mathrm{\bot}$ for various $\theta$ at around $\nu_\mathrm{B}$ = 4 at 28 mK. $R_{xy}$$-$$B$$_\mathrm{\bot}$ curves include an offset by 0.05 $h$/$e$$^{2}$ for each curve. ( b ), ( c ) $R_{xy}$ as a function of $B$$_\mathrm{//}$ at 28 mK for ($\nu_\mathrm{B}$, $\nu_\mathrm{G1}$, $\nu_\mathrm{G2}$) = (4, 2, 2). The arrows illustrated represent the spin orientation of the relevant edge channels in each $B$$_\mathrm{//}$ range.}
\end{figure*} 

Figure 3(a) displays $R_{xy}$ as a function of $V_\mathrm{G1}$ for $\nu_\mathrm{B} =$ 4 ($B =$ 2.01 T) and $\nu_\mathrm{G2} =$ 2 ($V_\mathrm{G2} =$ -1.10 V)  at 23 mK. When $V_\mathrm{G1}$ is reduced down to about -0.70 V, a deviation of $R_{xy}$ from 0.25 $h$/$e$$^{2}$ ($\Delta$$R_{xy}$) can be seen, and then $R_{xy}$ reaches 0.46 $h$/$e$$^{2}$ at $V_\mathrm{G1} =$ -1.10 V. Taking the relationship of $\Delta$$R_{xy}$ $=$ $\exp$($-$$L$$_\mathrm{edge}$/$L$$_\mathrm{eq}$)($h$/4$e$$^{2}$),\cite{Komiyama} where $L$$_\mathrm{eq}$ is the equilibration length corresponding to the distance over which electrons are traveling adiabatically, we roughly find $L$$_\mathrm{eq}$ $\approx$ 57 $\mu$m at 23 mK, being even larger than that of the high-mobility 2DEG in AlGaAs/GaAs heterostructures.\cite{Muller2} We also observe the evident temperature dependence of $R_{xy}$ as shown in Fig. 3(b), in which $\Delta$$R_{xy}$ decreases with increasing temperature. This means that the IEC scattering is accelerated and $L$$_\mathrm{eq}$ shrinks due to the increase in temperature. The results presented are the first experimental data associated with the edge-channel transport of high-mobility Si/SiGe heterostructures. 

In Fig. 4(a) we show the plots of $R_{xy}$ $vs$ $B$$_\mathrm{\bot}$ ($R_{xy}$$-$$B$$_\mathrm{\bot}$ curve) at around $\nu_\mathrm{B} =$ 4 for various $\theta$ in detail, where $B$$_\mathrm{\bot}$ $=$ $B$$_\mathrm{total}$ $\cos \theta$. In  66.5$^{\circ}$ $\lesssim$ $\theta$ $\lesssim$ 67.5$^{\circ}$, the plateau in the QH regime of $\nu_\mathrm{B} =$ 4 becomes unclear, which is general behavior of $R_{xy}$ under around coincidence condition.\cite{Zeitler2} Consequently, we can approximately regard the coincidence angle of the first Landau-level crossing of our sample as $\theta =$ 66.5$^{\circ}$. We also confirmed the coincidence in the vicinity of $\theta =$ 66.5$^{\circ}$ in $R_{xx}$$-$$B$$_\mathrm{\bot}$ curves. Assuming the effective mass $m$* $=$ 0.19 $m_\mathrm{0}$, where $m_\mathrm{0}$ is the free electron mass, we can deduce $g$* $=$ 4.2, being consistent with previous studies.\cite{Weitz,Schumacher,Koester,Lai2} At the coincidence angle ($\theta =$ 66.5$^{\circ}$), the spin orientations of the relevant edge channels are transferred from (0 $\downarrow$, 0 $\uparrow$) to (0 $\downarrow$, 1 $\downarrow$): the edge-channel transport in $\theta$ $\lesssim$ 66.0$^{\circ}$ or in $\theta$ $\gtrsim$ 68.0$^{\circ}$ arises from (0 $\downarrow$, 0 $\uparrow$) or (0 $\downarrow$, 1 $\downarrow$) edge channels, respectively.

To get insight into the spin dependence of the edge-channel transport in the Si/SiGe heterostructure, we examine $R_{xy}$ as a function of $\theta$ systematically at ($\nu_\mathrm{B}$, $\nu_\mathrm{G1}$, $\nu_\mathrm{G2}$) = (4, 2, 2), and summarize the dependence of $R_{xy}$ on $B$$_{//}$, where $B$$_{//}$ $=$ $B$$_\mathrm{total}$ $\sin \theta$, at 28 mK in Figs. 4(b) and (c). A deviation of $R_{xy}$ from 0.25 $h$/$e$$^{2}$ expresses suppression of the IEC scattering. We find that the value of $R_{xy}$ is nearly constant, i.e., 0.42 $h$/$e$$^{2}$ $\lesssim$ $R_{xy}$ $\lesssim$ 0.46 $h$/$e$$^{2}$, in $\theta$ $\lesssim$ 66.0$^{\circ}$ ($B$$_{//}$ $\lesssim$ 4.47 T) whereas $R_{xy}$ is markedly reduced at around the coincidence angle $\theta =$ 66.5$^{\circ}$ ($B$$_{//}$ $\simeq$ 4.59 T), and then the value of $R_{xy}$ is settled down to $R_{xy}$ $\sim$ 0.25 $h$/$e$$^{2}$ in $\theta$ $\gtrsim$ 71.5$^{\circ}$ ($B$$_{//}$ $\gtrsim$ 5.5 T). Komiyama {\it et al.} have reported that a spatial separation ($\Delta X$) between edge channels affects the IEC scattering for 2DEG in AlGaAs/GaAs heterostructures with $m$* $=$ 0.067 $m_\mathrm{0}$ and $g$* $=$ -0.44.\cite{Komiyama} In general, if $\Delta E_\mathrm{z}$ is enhanced by increasing $\theta$, $\Delta X$ between spin-resolved edge channels increases and the IEC scattering is suppressed due to the reduction in the overlap of electron wave functions.\cite{Komiyama} However, the above interpretation can not be applied to the data in Figs. 4(b) and (c).
\begin{figure}[t]
\includegraphics[width=8cm]{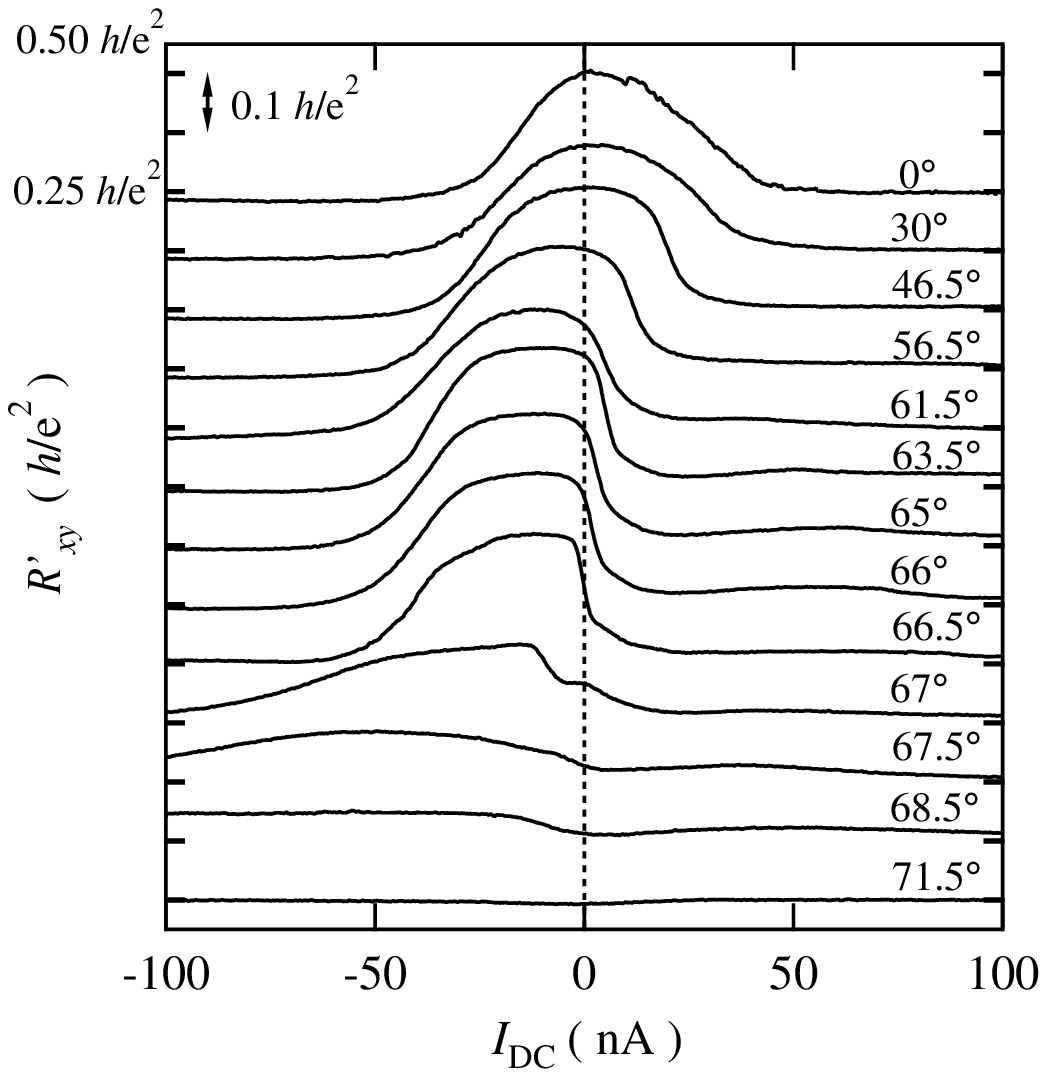}
\caption{$R'_{xy}$ $vs$ $I_\mathrm{dc}$ for different $\theta$ at 28 mK for ($\nu_\mathrm{B}$, $\nu_\mathrm{G1}$, $\nu_\mathrm{G2}$) = (4, 2, 2). The data traces of $R'_{xy}$ include an offset by 0.1 $h$/$e$$^{2}$ for each curve, and the major ticks are presented at every 0.1 $h$/$e$$^{2}$.}
\end{figure}

In order to explain the above feature, we attempt to approximately calculate $\Delta X$.\cite{Komiyama,Hirai,Muller2} Here, we use a parabolic-type confining potential with the confinement frequency of 1.7 $\times$ 10$^{12}$ s$^{-1}$, $m$* $=$ 0.19 $m_\mathrm{0}$, and $g$*$=$ 4.2. For $\theta =$ 0$^{\circ}$, $\Delta E_\mathrm{z}$ ($=$ $g$*$\mu_{B}B$, where $\mu_{B}$ is Bohr's magneton) of 0.4 meV indicates $\Delta X$ $\sim$ 47.5 $\AA$ at 2.01 T for the sample used. With increasing $\theta$, $\Delta E_\mathrm{z}$ is enlarged but the related $\Delta X$ is always smaller than 145 $\AA$ which is the maximum value of $\Delta X$ derived from the Landau gap ($\hbar$$\omega$$_\mathrm{c}$) of 1.22 meV.\cite{Komiyama,Hirai,Muller2} Since the magnetic length $l_{c} =$ $\sqrt{\hbar/(eB)}$ is $\sim$ 180 $\AA$, a strong mixing of the wave functions of electrons between edge channels can be deduced irrespective of $\theta$. Hence we conclude that the wave functions of electrons between edge channels usually overlap for the high-mobility Si/SiGe heterostructure used. This feature basically originates from the fact that $m$* of Si/SiGe heterostructures is large relative to that of AlGaAs/GaAs heterostructures by a factor of 3. We also note that the edge-channel transport is ascribed to the spin orientation of the relevant edge channels either (0 $\downarrow$, 0 $\uparrow$) or (0 $\downarrow$, 1 $\downarrow$): we can see the long $L$$_\mathrm{eq}$ in (0 $\downarrow$, 0 $\uparrow$) while a considerably shorten $L$$_\mathrm{eq}$ is found in (0 $\downarrow$, 1 $\downarrow$). 

For 2DEG in AlGaAs/GaAs heterostructures, M\"uller {\it et al.}\cite{Muller2} explained that $L$$_\mathrm{eq}$ of electrons in spin-resolved edge channels is inversely proportional to the spinor overlap, $|$$\chi_\mathrm{\downarrow}^{\dag}$($k_{i}$)$\chi_\mathrm{\uparrow}$($k_{f}$)$|$$^{2}$, where $i$ and $f$ denote the initial and final states in the scattering process of electrons. The spinor overlap can be written as $\chi_\mathrm{\downarrow}^{\dag}$($k_{i}$)$\chi_\mathrm{\uparrow}$($k_{f}$) $\propto$ ($g$*$\mu_{B}B$)$\gamma$$\hbar$$\delta k$/$\{$($g$*$\mu_{B}B$)$^{2}$ $+$ $\gamma$$^{2}$(2$\hbar k$)$^{2}$$\}$, where $\delta k$ $=$ $k_{f} - k_{i}$ and $\gamma$ is the spin-orbit coupling constant.\cite{Muller2,Khaetskii,Takagaki} They suggested that large values of $L$$_\mathrm{eq}$ $\sim$ 100 $\mu$m in spin-resolved edge channels can be interpreted by the small spinor overlap.\cite{Muller2,Khaetskii} We also obtain the long $L$$_\mathrm{eq}$ $\sim$ 57 $\mu$m between (0 $\downarrow$, 0 $\uparrow$) edge channels, implying the small spinor overlap, although the wave functions of electrons between edge channels are strongly mixed for the 2DEG in Si/SiGe heterostructure used, as mentioned in previous paragraph. In this regard, we infer that a small contribution of the spin-orbit interaction, derived from the inversion symmetry of a unit cell of Si crystal, causes the small spinor overlap of the above equation, and leads to suppression of the IEC scattering with spin-flips. On the other hand, we judge that the IEC scattering between (0 $\downarrow$, 1 $\downarrow$) edge channels without spin-flips is accelerated due to the overlap of the wave functions of electrons. Although the effect of the hyperfine interaction between electron and nuclear spins is also predicted, we can rule out it because 95.33 \% of nuclear isotopes ($^{28}$Si and $^{30}$Si) in Si has no nuclear moment.

We finally refer to the IEC scattering controlled by $I_\mathrm{dc}$. At ($\nu_\mathrm{B}$, $\nu_\mathrm{G1}$, $\nu_\mathrm{G2}$) = (4, 2, 2), when the positive direct current, $I_\mathrm{dc}$ $>$ 0 ($\mu_\mathrm{S}$ $>$ $\mu_\mathrm{D}$), is applied between inner ($\mu_\mathrm{D}$) and outer  ($\mu_\mathrm{S}$) edge channels, the IEC scattering from outer to inner occurs markedly, while the IEC scattering from inner to outer becomes significant in $I_\mathrm{dc}$ $<$ 0 ($\mu_\mathrm{S}$ $<$ $\mu_\mathrm{D}$). Thus, the differential Hall resistance, $R'_{xy} =$ $\partial V$$_{xy}$/$\partial I$, as a function of $I_\mathrm{dc}$ ($R'_{xy}$$-$$I_\mathrm{dc}$ curve) shows characteristic nonlinearity.\cite{Komiyama,Komiyama3,Muller3,Machida} Figure 5 shows $R'_{xy}$$-$$I_\mathrm{dc}$ curve for various applied magnetic field directions $\theta$. For $\theta =$ 0$^{\circ}$, a marked nonlinear feature is seen in $I_\mathrm{dc} \lesssim$ $\pm$ 40 nA and the symmetry of the $R'_{xy}$$-$$I_\mathrm{dc}$ curve is comparatively maintained in that regime. In contrast, the IEC scattering is promoted and $R'_{xy}$ becomes 0.25 $h$/$e$$^{2}$ in $I_\mathrm{dc} \gtrsim$ $\pm$ 50 nA. With $\theta$ increased, the symmetric shape of the $R'_{xy}$$-$$I_\mathrm{dc}$ curve is broken and the shift of the nonlinear region toward $I_\mathrm{dc}$ $<$ 0 is observed. For 2DEG in AlGaAs/GaAs heterostructures, nonlinear features shown in $R'_{xy}$$-$$I_\mathrm{dc}$ curves are explained by the rearrangement of edge channels due to unequal edge-channel population.\cite{Komiyama,Komiyama3,Muller3,Machida} On the other hand, for 2DEG in the Si/SiGe heterostructure we use, the above $I_\mathrm{dc}$ dependence of $R'_{xy}$ can not be interpreted by this explanation. The cause of this asymmetric feature is still unclear but the $R'_{xy}$ $-$ $I_\mathrm{dc}$ curves vary systematically with increasing $\theta$ under around the coincidence condition, strongly supporting that these $I_\mathrm{dc}$ dependence of $R'_{xy}$ are associated with the spin dependence of the edge-channel transport described. Therefore, this should be considered to be a peculiar property of the 2DEG in Si/SiGe heterostructures. 

In summary, we have studied the edge-channel transport in the high-mobility 2DEG in a Si/SiGe heterostructure in the QH regime. We observed the spin-dependent edge-channel transport at around the Landau-level crossing in tilted magnetic fields. The evident spin dependence is due to a small contribution of the spin-orbit interaction in Si to the spin-flip IEC scattering.

K. H. and T. M. acknowledge Prof. S. Komiyama, Dr. M. Jung and Dr. A. Umeno of University of Tokyo for many useful discussion and experimental supports. This work is partially supported by the Inamori foundation and the Support Center for Advanced Telecommunications Technology Research.



\begin{thebibliography}{11}
\bibitem{ref}
{The adiabatic edge-channel transport could not be seen at ($\nu_\mathrm{B}$, $\nu_\mathrm{G1}$, $\nu_\mathrm{G2}$) = (2, 1, 1) in our device. Hence, IEC scattering was predominant between the valleys: we regarded the valleys as a degenerate channel on the edge-channel transport in this study.}
\end{thebibliography}
\end{document}